\documentstyle[aps,prb,preprint] {revtex}
\begin{document}
\draft
\preprint{}
\title{Depletion of density of states near Fermi energy induced by
disorder and electron correlation in alloys}
\author{Han-Jin Noh}
\address{Department of Physics, Seoul National University, Seoul
151-742, South Korea}
\author{Tschang-Uh Nahm \cite{tun}}
\address{Department of Physics, Hanyang University, Seoul 133-791,
South Korea}
\author{Jae-Young Kim, W.-G. Park,\cite{wgp} and S.-J. Oh}
\address{Department of Physics, Seoul National University, Seoul
151-742, South Korea}
\author{J.-P. Hong and C.-O. Kim}
\address{Department of Physics, Hanyang University, Seoul 133-791,
South Korea}
\date{\today}
\maketitle

\begin{abstract}
We have performed high resolution photoemission study of
substitutionally disordered alloys Cu-Pt, Cu-Pd, Cu-Ni, and Pd-Pt.
The ratios between alloy spectra and pure metal spectra are found to
have dips at the Fermi level when the residual resistivity is high and when
rather strong repulsive electron-electron interaction is expected.  This is
in accordance with Altshuler and Aronov's model which predicts depletion of
density of states at the Fermi level when both disorder and electron
correlation are present.
\end{abstract}


\narrowtext

\section{Introduction}
\label{sec:1}

Two decades ago Altshuler and Aronov interpreted tunneling data of some
disordered systems as showing the electron-electron interaction effect which
induces change in the density of states (DOS) around Fermi level. \cite{alt79}
Their model predicted the reduction of DOS in the case of 3 dimensional solids
given by the relation
\[
\delta\nu = \frac{\lambda}{4\sqrt{2} \pi^{2}} \frac{\sqrt{|{\epsilon}|}}{(\hbar
D)^{3/2}},
\]
where $\delta\nu$ is change of the density of states, $\epsilon$ energy
measured from Fermi level, $D$ diffusion coefficient which is inversely
proportional to the disorder driven resistivity $\rho$, $\lambda$ the
electron-electron interaction strength. Recently it was applied to the
interpretation of the photoemission spectra of La(NiMn)O${_3}$ and
La(NiFe)O${_3}$ which undergo metal-insulator transition (MIT).\cite{sar98} The
reduction of DOS at the Fermi energy ($E_F$) was observed as Mn or Fe ions were
substituted, and it was interpreted as due to the disorder and electron
correlation effect following Altshuler and Aronov's theory. However, the
reduction of DOS has not been observed before for typical disorder metallic
alloy system, and this is the subject of present investigation.

Among metallic alloys, the most probable candidates showing both disorder and
electron correlation effect are the transition metal alloys having 9 $d$
electrons.  In fact, it is well known that the residual resistivity of many
disordered alloy systems such as Cu-Au and Pd-Pt show simple Nordheim behavior,
but most of binary alloys one component of which is noble-metal and the other
transition metal with approximately 9 $d$ valence electrons Ni, Pd, and Pt
(henceforth designated as $d$-9 transition metals) show great deviation from
the Nordheim behavior.\cite{mot36} For example, Pt-rich Cu-Pt and Pd-rich Ag-Pd
alloys have ill-defined Fermi surface especially around $X$ point due to the
broadening of the Bloch spectral function at and near the Fermi vector {\bf
k$_F$} \cite{pin80,ban89} and as a result, both theoretically and
experimentally an asymmetric residual resistivity curve with values of as high
as 85~$\mu \Omega$ cm has been observed. It is thus expected that due to strong
disorder at $E_F$ in noble - $d$-9 transition metal alloys and due to the
electron-electron interaction between Pt, Pd, or Ni ${d}$ electrons at the
Fermi level,\cite{mar80-1,mar80-2} one may observe the reduction of DOS at
$E_F$, although the dip size might be smaller than in Ref. \onlinecite{sar98}
due to screened Coulomb interaction between valence electrons.

In order to see such effect, we compared the angle-integrated photoemission
spectra of Cu-Pt, Cu-Pd, Cu-Ni, and Pt-Pd alloys near $E_F$ with those of pure
$d$-9 transition metals which have almost smooth DOS near $E_F$.  We observed
some depletion of DOS at $E_F$ in Cu$_{25}$Pt$_{75}$ and possibly for
Cu$_{50}$Pt$_{50}$ and Cu$_{25}$Ni$_{75}$, the residual resistivity values of
which are almost 3 times higher than those of high Cu concentration alloys due
to strong disorder effect at some portion of the Fermi surface which reduces
electron mean free path.  This along with the absence of depletion for
Pd$_{50}$Pt$_{50}$ whose electron mean free path is expected to be much longer
due to sharp Bloch spectral function at $E_F$ (Ref.\onlinecite{flo89}) show
clear signature that the Altshuler and Aronov's theory is applicable to {\em
ordinary} disordered transition metal alloys as well as to transition metal
oxides showing MIT.\cite{sar98}

\section{Experimental}
\label{sec:2}

The polycrystalline Cu-Pt, Cu-Pd, Cu-Ni, and Pd-Pt alloys were made by arc
melting of 99.99\% Cu, 99.9\% Pt, 99.95\% Pd, and 99.997\% Ni, and then
annealed at temperature of 950$^{\circ}$C to ensure disorder phase which was
confirmed with x-ray diffraction. The data were taken with VG Microtech CLAM 4
multi-channeltron detector electron energy analyzer with resolution of 40~meV
full width at half maximum (FWHM), and the base pressure was $1.5 \times
10^{-10}$ torr. Photon source was unmonochromatized He~{\footnotesize I} line
($h\nu$ = 21.2 eV).  The samples were cooled down to 95~K with liquid nitrogen
and the surfaces were scraped {\em in situ} at that temperature.  In order to
prevent contamination during the experiment, further scraping was done
frequently.

\section{Results and Discussion}
\label{sec:3}

Figure~1 shows the photoemission spectra of disordered Cu$_{25}$Pt$_{75}$ along
with that of Pt near $E_F$.  We can see that the lineshape of
Cu$_{25}$Pt$_{75}$ and that of Pt around $E_F$ are not quite the same, which
indicates the detailed difference in DOS.  Since the difference of spectral
lineshape between Cu$_{25}$Pt$_{75}$ and Pt is quite small, we tested several
possibilities which may induce such change besides the difference of DOS.
Apparently, this kind of experimental lineshape difference cannot be attributed
to a result of Fermi-Dirac distribution functions at slightly different
temperatures.  If it really were, the spectrum of a specimen at the lower
temperature should have smaller spectral intensity at $E > E_F$ and more at $E
< E_F$ and this difference should be symmetric about $E_F$, which is not the
case in the figure.  We even tried to fit the alloy spectrum with broadened Pt
spectra to take account of possible temperature difference, but the lineshape
was not the same at all. Repeating the measurements with new batch of samples
also confirmed this behavior.  Also, one might think that the Fermi level
alignment can cause much uncertainty on the comparison of the two spectra, but
when we shifted one of the spectra being compared by more than 2~meV, which was
beyond the experimental error, the alignment was completely out of sense.

In order to see whether there really is some difference in the DOS near the
Fermi level, we divided the spectrum of the alloy by that of pure Pt, assuming
that the Pt DOS near $E_F$ does not have any sharp structure.  The absence of
sharp structures in pure Pt DOS is confirmed by fitting the Pt spectrum with a
linearly changing DOS, which is reasonably a good approximation within the
energy region of interest. \cite{mat80}  This approximated DOS was multiplied
by the Fermi-Dirac distribution and convoluted with a Gaussian curve
corresponding to the instrumental resolution.  The fitting was found to be
perfect as expected, and the same was true for pure Cu spectrum.

The divided spectrum clearly reveals the presence of a dip at $E_F$ which
extends to almost 50~meV above and below the Fermi level.  The dip size is very
small with its relative value of only 6~\% of the divided spectral weight near
the ends of the dip.  The slope in the divided spectra seems to be caused by
the detailed band structures of alloys and of pure Pt.

This dip structure could not be removed by any of the assumptions such as the
mis-alignment of the Fermi level, temperature difference, inclusion of the
satellites of He~{\footnotesize I} line and the background treatment. These
factors have their own effects when we simulate a curve, but none of them could
be an explanation of the dip.  For instance, the smooth decrease or increase of
the divided spectral weight above the Fermi level was observed when there is
slight difference in the temperature.  If the temperature at which the pure Pt
spectrum was taken is higher by only 1~K, great decrease of the divided spectra
above $E_F$ was observed.  However, the presence of a dip near $E_F$ was not
disputable since the shape of the divided spectrum below $E_F$ did not change
much with the temperature difference of about 1~K. We also checked the
possibility of mis-alignment of the Fermi level as a probable explanation for
the dip, but it was found to give only smooth change in the divided spectrum,
not affecting the presence of a dip.

One might think that the depletion of DOS at the Fermi level could be
understood as a real structure in calculated DOS in one electron picture.
However, the coherent-potential-approximation (CPA) calculation which gives
good description of one-electron DOS of alloys does not show any hint of a
sharp dip at the Fermi level.\cite{ban89-2}  It is also well known that when
$d$-9 transition metals form alloys with other metals, the DOS near $E_F$ is
lowered in some cases,\cite{fug83} but this happens only for low concentration
of $d$-9 transition metals, not for high concentration. Furthermore, any sharp
structure near $E_F$, if existed, should manifest itself more prominently in
ordered phase rather than in disordered phase because of strong broadening of
Bloch spectral function due to disorder for the latter. To the best of our
knowledge, there have not been any report about a presence of sharp structure
in high resolution photoemission spectra of pure metals. Therefore, this kind
of reduction of DOS at the Fermi level we believe can only be interpreted as a
result of the electron-electron interaction in disordered system.

In Fig.~2, we show divided spectra of Cu-Pt alloys with Pt concentration of 25,
50, and 75 at.\% to see the possible sharp change of DOS near $E_F$ as a
function of composition. From the figure, it is apparent that any such
structure at $E_F$ does not exist in the case of Cu-rich alloy, and it can be
safely argued that dips at $E_F$ for Cu$_{25}$Pt$_{75}$ and Cu$_{50}$Pt$_{50}$
cannot be understood within one-electron picture.  For Cu-Pt alloys, the
residual resistivity has its maximum value at 60 at.\% Pt, and the reduction of
DOS near $E_F$ should be the greatest around that concentration if we assume
constant $\lambda$ throughout the whole concentration.  This indeed is in
agreement with the experiment where the depletion of DOS is observed only for
Cu$_{25}$Pt$_{75}$ and Cu$_{50}$Pt$_{50}$, the latter having a dip with only
3~\% reduction.

To observe similar effect, we have also studied Cu-Ni alloys. It is generally
believed that the electron-electron interaction between $d$ electrons is
stronger for Ni 3$d$ than for Pt 5$d$ because of more localized wavefunctions.
In fact, from the spectra of Cu$_{25}$Ni$_{75}$ and of Ni and the divided
spectrum shown in the upper panel of Fig.~3, the reduction of DOS is clearly
visible.  On the other hand, the divided spectrum of Cu$_{50}$Ni$_{50}$ in the
lower panel did not show such behavior. Although the values of $\rho_0$ are
almost the same for these two compositions, the electron correlation effect of
Ni 3$d$ must be small for Cu$_{50}$Ni$_{50}$ than for Cu$_{25}$Ni$_{75}$
because it is known that the partial DOS of Ni 3$d$ at $E_F$ diminishes as Ni
is diluted. \cite{sto71} Therefore the absence of a dip for Cu$_{50}$Ni$_{50}$
alloy can be explained within the context of Altshuler and Aronov's model.

We also investigated the spectra of Cu-Pd alloys with 25, 50, 75, 90 at.\% of
Pd. In this case, we could not see any reduction of DOS as shown for the
representative Cu$_{25}$Pd$_{75}$ case in the upper panel of Fig.~4. This
probably is due to smaller residual resistivity $\rho_0$ than Cu-Pt alloys and
weaker electron-electron interaction than in Cu-Ni alloys, and a dip at $E
\simeq$ $-$~50~meV seems to be a result of deviation of Pd DOS itself near
$E_F$ from almost linearly decreasing behavior within the energy region
studied.

The absence of a dip for Cu-Pd alloys is actually against another possible
explanation of depletion of DOS near $E_F$, namely the disorder induced
non-uniform lifetime broadening of Bloch function.  If the broadening is
present only around the Fermi level, one may expect the depletion of DOS at the
Fermi level because some of the spectral weight will be piled up at lower and
higher energy side.  Were it true, this kind of depletion should be observed in
Cu-Pd alloys as well as in Cu-Pt and Cu-Ni alloys, because the broadening of
Bloch spectral function near {\bf k$_F$} must be more or less similar to Cu-Ni
and Cu-Pt alloys when $d$-9 transition metal concentration is between 50 to
80\%.  In fact, the similarity of Bloch spectral functions can be easily seen
in the CPA calculation results of Ag-Pd,\cite{pin80} of Cu-Pt,\cite{ban89} and
of Cu-Ni,\cite{dur80} which is expected on the ground that they have almost the
same composition dependence of $\rho_0$.  However, we do not see any prominent
structure at all for Cu-Pd alloys, in contrast to this scenario.

The ill-defined Fermi surface itself does not mean explicitly that the electron
at $E_F$ has relatively small mean free path, because the electronic states
with short lifetime slightly below or above the Fermi level can cause it. Pd-Pt
alloys show such behavior.  In this case, although the Fermi surface could not
be defined clearly because of short lifetime of the electrons just below and
above the Fermi level, the spectral function becomes very sharp as the
wavevector {\bf k} crosses {\bf k$_F$}.\cite{flo89}  This results in an
ordinary Nordheim behavior of $\rho_0$.  If such non-uniform lifetime
broadening would cause some change in DOS, we may expect some increase of DOS
at $E_F$ and small decrease just above and below $E_F$ in the experimental
spectra. However, as in the lower panel of Fig.~4, we could not see such
structure for Pd$_{50}$Pt$_{50}$ with small $\rho_0$ value.  This shows that
both disorder and electron correlation effect are necessary to observe a dip at
$E_F$ in agreement with Altshuler and Aronov's model. For Cu-Pt alloys, the
spectral function does have finite lifetime even at {\bf k$_F$} which accounts
for the relatively high $\rho_0$ values and this is the reason we could observe
disorder and electron interaction induced depletion of DOS.

The existence of the electron-electron interaction for the Cu alloys with
different $d$-9 transition metals observed here suggests important implication
for the understanding of their electronic structures . The transition metals
with 9 $d$ electrons in their outermost shell have strong correlation effect
and it is now well known that the so-called 6~eV satellite in photoemission
spectrum of Ni (Ref.\onlinecite{mar80-1}) can only be understood with the
inclusion of many-body interaction.  For Pd, it was suggested that the
satellites of core levels and even of the valence band must be due to the same
reason.\cite{mar80-2} From our results, it is quite likely that even pure Pt
metal can have some electron correlation effect although this interaction might
be largely screened by the electrons at {\bf k${_F}$} the band of which is more
dispersive than of Ni 3$d$.  In fact, the presence of satellite structure in Pt
4$f$ core level spectra suggests rather strong electron-electron
interaction.\cite{nah00}

\section{Conclusion}
\label{sec:4}

To conclude, we have studied the photoemission spectra of transition metal
alloys and pure elements.  There were dips at $E_F$ for some alloys with high
$\rho_0$ values when one of the two elements have non-negligible electron
correlation effect at $E_F$.  The dip size at $E_F$ and its width in this work
were smaller than those observed for a material with pseudo-gaps \cite{sus99}
and much smaller than observed near MIT of transition metal
compounds,\cite{sar98} but their existence was clearly established.  We have
argued that substitutional disorder in alloys cause depletion of DOS at $E_F$
when the electron-electron interaction is present, and this could be
qualitatively interpreted within the context of Altshuler and Aronov's
model.\cite{alt79}

\acknowledgements T.-U. N. wishes to acknowledge the financial support of
Hanyang University, Korea, made in the program year of 1998.  This work was
also supported in part by the Korean Science and Engineering Foundation through
Center for Strongly Correlated Materials Research (CSCMR) at Seoul National
University.

\begin{figure}
\caption {Photoemission spectra of Cu$_{25}$Pt$_{75}$ (line) and of pure Pt
(open squares) as well as the divided spectra I(Cu$_{25}$Pt$_{75}$)~/~I(Pt)
(solid circles) with $h\nu$ = 21.2~eV at 95~K.  The spectra were normalized
after their backgrounds were removed.}
\end{figure}

\begin{figure}
\caption {Divided spectra I(Cu$_{x}$Pt$_{1-x}$)~/~I(Pt) for x = 0.25, 0.50, and
0.75.  Note the absence of a dip for x = 0.75.}
\end{figure}

\begin{figure}
\caption {Photoemission spectra of alloys, pure elements, and the divided
spectra for Cu$_{25}$Ni$_{75}$ (upper panel) and Cu$_{50}$Ni$_{50}$ (lower
panel). Note the presence of a dip at the Fermi level for Cu$_{25}$Ni$_{75}$.}
\end{figure}

\begin{figure}
\caption{Photoemission spectra of alloys, pure elements, and the divided
spectra for Cu$_{25}$Pd$_{75}$ (upper panel) and Pd$_{50}$Pt$_{50}$ (lower
panel).}
\end{figure}

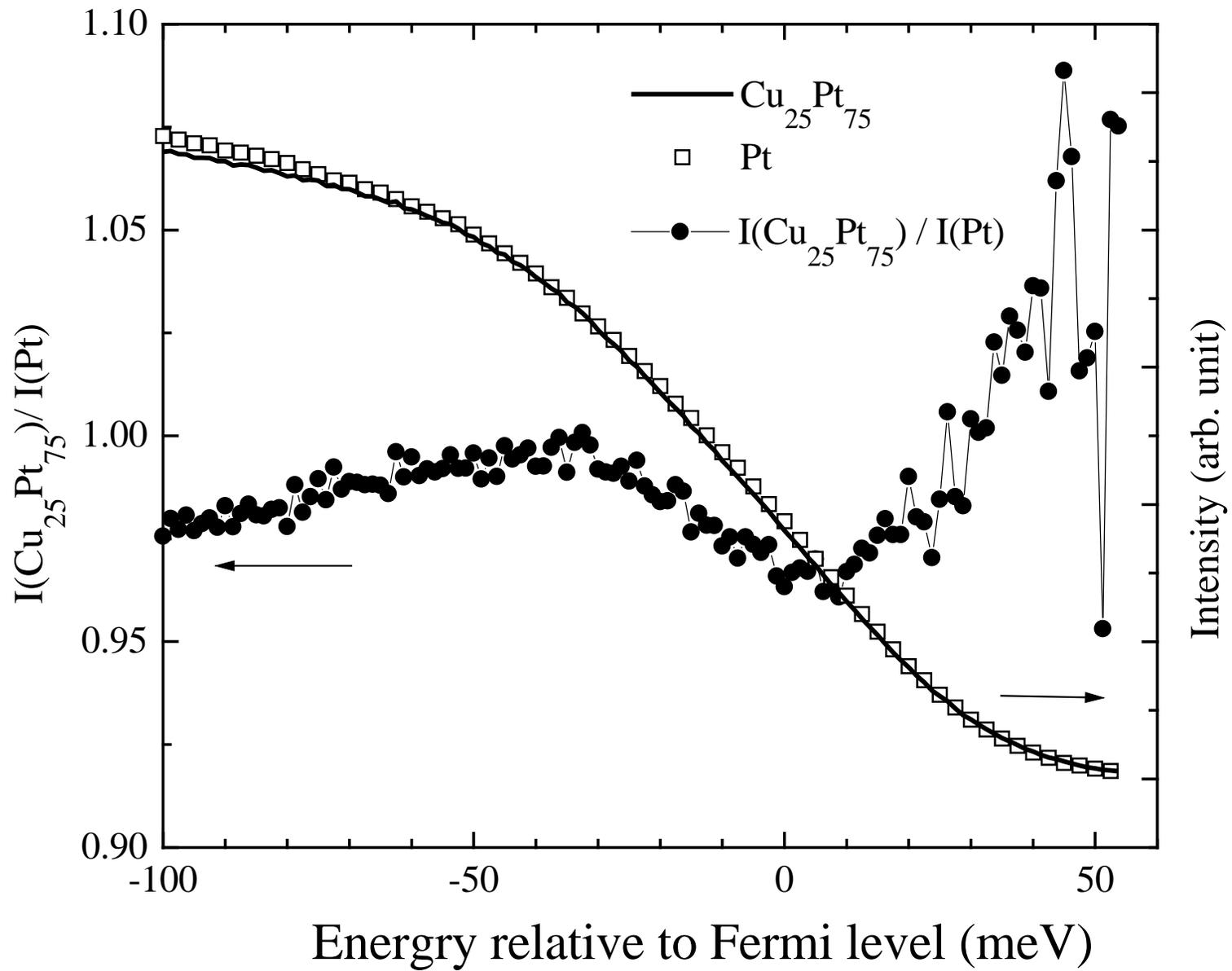
\end{document}